\documentclass[aps,twocolumn,superscriptaddress,amsfont,graphicx,preprintnumbers]{revtex4-1}
\renewcommand{\section}[1]{\textbf{\textit{#1.---\!}}}

\usepackage[utf8]{inputenc}

\usepackage[pdftex]{graphicx}
\usepackage[colorlinks=true
,urlcolor=blue
,anchorcolor=blue
,citecolor=blue
,filecolor=blue
,linkcolor=blue
,menucolor=blue
,pagecolor=blue
]{hyperref}
\usepackage[all]{hypcap}

\pdfoutput=1
\usepackage{bbm}
\usepackage{amsmath,amssymb}
\usepackage[usenames,dvipsnames]{color}
\usepackage{slashed} 

\makeatletter
\def\endfmffile{%
  \fmfcmd{\p@rcent\space the end.^^J%
          end.^^J%
          endinput;}%
  \if@fmfio
    \immediate\closeout\@outfmf
  \fi
  \IfFileExists{\thefmffile.mp}{\immediate\write18{mpost \thefmffile}}{}
  \let\thefmffile\relax
}
\makeatother
\usepackage[normalem]{ulem}

\newcommand{\cf}{{\it c.f. }}
\newcommand{\ie} {{\it i.e. }}

\newcommand {\beq} {\begin{equation}}
\newcommand {\eeq} {\end{equation}}
\newcommand {\bea} {\begin{eqnarray}}
\newcommand {\eea} {\end{eqnarray}}
\newcommand{\comment}[1]{}

\allowdisplaybreaks

\usepackage{youngtab}
\usepackage{multirow}

\newcommand{\SU}{\text{SU}}
\newcommand{\U}{\text{U}}
\newcommand{\Sp}{\text{Sp}}

\newcommand{\refeq}[1]{Eq.(\ref{#1})}

\begin{document}

\title{A composite scalar hint from di-boson resonances?}
\date{\today}

\author{Haiying Cai}
\affiliation{Universit\'e de Lyon, France; Universit\'e Lyon 1, Villeurbanne, France; \\ CNRS/IN2P3, UMR5822, IPNL F-69622 Villeurbanne Cedex, France}
\author{Thomas Flacke}
\affiliation{Department of Physics, Korea University, Seoul 136-713, Korea}
\author{Mickael Lespinasse}
\affiliation{Universit\'e de Lyon, France; Universit\'e Lyon 1, Villeurbanne, France; \\ CNRS/IN2P3, UMR5822, IPNL F-69622 Villeurbanne Cedex, France}

%
%

\begin{abstract}
We study the light scalar resonance sector of a composite Higgs model UV embedding based on the coset $SU(4)/Sp(4)$. Beyond the Higgs multiplet, the pseudo Nambu-Goldstone sector of this model contains Standard Model singlets which couple to the Standard Model gauge bosons through Wess-Zumino-Witten anomaly terms. They can thus be produced in gluon fusion and decay into either gluons or pairs of electroweak gauge bosons $WW$, $ZZ$, $Z\gamma$, or $\gamma\gamma$. In this letter we show that one of the pseudo Nambu-Goldstone boson states has appropriate couplings in order to explain a di-boson excess in the $WW$ channel whilst not being excluded by LHC run I bounds on the di-jet, $Z\gamma$ and $\gamma\gamma$ decay channels. A di-boson resonance production cross section of $\sim$ 10 fb at LHC run I is not a prediction of the model, but can be obtained if the confining gauge group is of high rank.   
\end{abstract}

\maketitle



\section{Introduction}

In this letter we consider a concrete example of a composite Higgs model UV embedding and show that it contains a pseudo Nambu Goldstone boson (pNGB) which yields  di-boson resonance signatures. The model is based on  a confining gauge theory with 4 fermions $Q$ and 6 fermions $\chi$.  The former provide an $SU(4)$ flavor symmetry which -- when spontaneously broken into $Sp(4)$ -- provide a composite Higgs (and two SM singlets) as pNGBs. The $\chi$ fermions in a $\bf{6}$ of $SU(6)$ (spontanously or explicitly broken into $SO(6)$) is required in order to obtain colored fermionic resonances which can serve as top partners. The model is proposed in Ref.~\cite{Barnard:2013zea}, where the $\chi$ fermions are in a $SU(3)_1 \times SU(3)_2$ global symmetry,  with the aim to construct a UV embedding of a composite Higgs model which does not contain any elementary scalars. Ref.~\cite{Cacciapaglia:2015eqa} extends the global symmetry to be $SU(4) \times SU(6)$ and has studied in particular the pNGB sector arising from the $SU(6)/SO(6)$ breaking. Apart from a colored sextet and octet, the pNGB sector from $\chi \chi$ condensation contains one further SM singlet. Those SM singlets from the breaking of  $SU(4) \times SU(6)$ symmetry  thus arise as a  by-product in the construction of a UV embedding. Obtaining additional pNGBs beyond the Higgs multiplet is common in composite Higgs UV embeddings and actually a necessity when demanding no elementary scalars in combination with a composite Higgs pNGB multiplet and the existence of top partners with the correct quantum numbers (\cf Refs.~\cite{Ferretti:2013kya, Vecchi:2015fma}). 
In this letter we show that in the particular case of  $SU(4) \times SU(6)/ (Sp(4) \times SO(6))$ model, one linear combination of the singlet pNGBs is a pseudo scalar with anomaly couplings as envisioned in Ref.~\cite{Cacciapaglia:2015nga}. 

Our initial investigation presented here shows that this resonance can yield an explanation for the recently reported di-boson excess if its mass is chosen to 2 TeV.  
In Ref.~\cite{Aad:2015owa}, ATLAS reported a  $3\sigma$ excess in the hadronic $WW,WZ,ZZ$ resonance channel for a narrow resonance at around 2 TeV with a cross section of $\sim 10$ fb. The corresponding CMS search \cite{Khachatryan:2014hpa} found a less pronounced excess at the same mass range. At the same time, no significant excess is found in this mass range in searches of semi-leptonic channels ~\cite{Khachatryan:2014gha,Aad:2015ufa} For a recent combination of the ATLAS and CMS results on di-boson searches \cf Ref.~\cite{Dias:2015mhm}. As pointed out in Ref.~\cite{Cacciapaglia:2015nga}, the diboson anomaly could be explained by a Standard Model neutral pseudo-scalar particle $\sigma$ which couples to the Standard Model (SM) exclusively through Wess-Zumino-Witten anomaly terms \footnote{Soon after, pseudo scalar di-boson candidates were also parameterized and discussed in Refs.\cite{Kim:2015vba,Fichet:2015yia}}. In such a scenario, $\sigma$ can be produced via gluon fusion and decay into $WW$ or $ZZ$, while the apparent resonance in the hadronic $WZ$ channel would need to arise from contamination by the $WW$ and $ZZ$. This assumption is supported by the analysis in Ref.~\cite{Allanach:2015hba}. Unlike vector resonances, a pseudoscalar resonance can also decay into $Z\gamma$ and $\gamma\gamma$ final states, and if the couplings arise solely through anomaly terms, the branching ratios into electroweak gauge bosons are fully fixed in terms of the quantum numbers of the underlying model, rendering this setup highly predictive and testable.  As pointed out in Ref.~\cite{Cacciapaglia:2015nga}, a  natural candidate of a pseudo scalar with anomalous couplings would be a composite state of a strongly coupled sector for which we here present an explicit example.

\bigskip

\section{A composite Higgs model with a scalar di-boson candidate}

We consider a composite Higgs model outlined in Refs. \cite{Barnard:2013zea,Cacciapaglia:2015eqa}. It is based on  a confining $Sp(2N_c)$ gauge theory with 4 fermions $Q$ in the fundamental and 6 fermions $\chi$ in the antisymmetric representation (c.f Table \ref{tab:fund_field_content} for the field content, the global, and SM gauge charges of the underlying fermions). The model exhibits an $SU(4)\times SU(6)\times U(1)$  global symmetry which is spontaneously broken to $Sp(4)\times SO(6)$ if chiral condensates $\langle QQ \rangle$ and $\langle \chi \chi \rangle$ form. It incorporates a pNGB Higgs as well as candidates for top partners amongst the three-fermion bound states and provides promising first steps towards an all-fermionic  UV embedded composite Higgs model.

\begin{table}[tb]
\begin{center}
\begin{tabular}{|c|c|c|c|c|c|c|c|}
\hline
& $\Sp(2N_c)$&${\SU(3)}_c$&${\SU(2)}_L$&${\U(1)}_Y$ & SU(4) & SU(6) & U(1) \\
\hline
$ \begin{array}{c} Q_1 \\ Q_2 \end{array} $&${\tiny{\yng(1)}}$&$\bf 1$&$\bf 2$&$0$& \multirow{3}{*}{\bf 4} & \multirow{3}{*}{\bf 1} & \multirow{3}{*}{$q_Q$}\\
\cline{1-5}
$Q_3$&${\tiny{\yng(1)}}$&$\bf 1$&$\bf 1$&$1/2$ & & &\\
\cline{1-5}
$Q_4$&${\tiny{\yng(1)}}$&$\bf 1$&$\bf 1$&$-1/2$ & & & \\
\hline
$ \begin{array}{c} \chi_1 \\ \chi_2 \\ \chi_3\end{array} $&${\tiny{\yng(1,1)}}$&$\bf 3$&$\bf 1$&$2/3$ & \multirow{4}{*}{\bf 1} & \multirow{4}{*}{\bf 6} & \multirow{4}{*}{$q_\chi$}\\
\cline{1-5}
$ \begin{array}{c} \chi_4 \\ \chi_5 \\ \chi_6\end{array} $&${\tiny{\yng(1,1)}}$&$\bf \bar{3}$&$\bf 1$&$-2/3$ & & &\\
\hline
\end{tabular} 
\caption{Field content of the microscopic fundamental theory and property transformation under the gauged symmetry group Sp(2$N_c$)$\times$SU(3)$_c \times$ SU(2)$_L \times$ U(1)$_Y$, and under the global symmetries SU(4)$\times$SU(6)$\times$U(1).}
\label{tab:fund_field_content}
\end{center}
\end{table}

The mesons $QQ$ and $\chi\chi$ each contain one $Sp(4)\times SO(6)$ singlet, $\sigma_Q$ and $\sigma_\chi$, which are associated to the spontaneously broken $U(1)_Q$ and $U(1)_\chi$. Only one linear combination of these -- which we explicitly determine below --  is $Sp(2N_c)$ anomaly free, leaving only one pNGB from this sector. The the other one is expected to obtain a large mass from $Sp(2N_c)$ instanton effects. In addition, $QQ$ contains a boson multiplet in the $(5,1)$ under $Sp(4)\times SO(6)$. In terms of the $SU(2)_L\times SU(2)_R\subset Sp(4)$, it decomposes into $(H,\eta)$ in $(2,2)\oplus(1,1)$, where the bi-doublet is identified with the SM-like Higgs while $\eta$ is another SM singlet pNGB.  

The states $\sigma_Q$, $\sigma_\chi$ and $\eta$ couple to pairs of SM gauge bosons  through anomalies. We parameterize the interactions as 
\beq
\mathcal{L}_{\sigma_i \mathcal{G}\mathcal{G}}=\frac{g^2_\mathcal{G}}{32\pi^2}\frac{\kappa^{i}_\mathcal{G} \sigma_i}{f_i}\epsilon^{\mu\nu\rho\sigma}\mathcal{G}^k_{\mu\nu}\mathcal{G}^k_{\rho\sigma}\, ,\\
\eeq
where $\sigma_i=(\sigma_Q,\sigma_\chi,\eta)$, $f_i$ are their decay constants, and $\mathcal{G}$ labels the couplings and field strengths of the gauge groups $SU(3)_c$, $SU(2)_L$ , and $U(1)_y$. The anomaly coupling coefficients $\kappa^i_\mathcal{G}$ are shown in Table \ref{tab:1}.

\begin{table}[tb]
\begin{center}
\begin{tabular}{c|c|c|c}
                       & $\sigma_Q$ & $\sigma_\chi$ & $\eta$ \\\hline
 $\kappa_g$   &            0             &   $(2N_c+1)(N_c-1) $ & 0 \\
 $\kappa_W$  &    $2 N_c $        &                  0                 & $2 N_c  \frac{\cos(v/f)}{2\sqrt{2}}$\\
 $\kappa_B$   &  $2N_c $            &  $ \frac{8}{3}(2N_c+1)(N_c-1)$ & $-2 N_c \frac{\cos(v/f)}{2\sqrt{2}}$ 
\end{tabular}
\end{center}
\caption{Anomaly coefficients of $\sigma_Q,\sigma_\chi$, and $\eta$.}
\label{tab:1}
\end{table}

The $Sp(2N_c)$ anomaly breaks $U(1)_Q\times U(1)_\chi\rightarrow U(1)_\sigma$. To identify the anomalous state and the anomaly-free pNGB, we start from the  Goldstone Lagrangian of  $\sigma_Q$ and $\sigma_\chi$,
\beq 
\mathcal{L}_{\rm kin,GB}=\frac{{{f^2_Q}}}{2}{\partial _\mu }\Sigma _{QQ}^{\dag}{\partial ^\mu }{\Sigma _{QQ}} + \frac{{{f_\chi^2}}}{2}{\partial _\mu }\Sigma _{\chi \chi }^{\dag} {\partial ^\mu }{\Sigma _{\chi \chi }},
\label{eq:L1}
\eeq
where
\beq
{\Sigma _{QQ}} = {e^{i \sigma_Q /f_Q}},  \quad  {\Sigma _{\chi \chi }} = {e^{i{ \sigma_\chi}/{f_\chi}}}.
\eeq
The conserved current (up to the anomaly) of a $U(1)$ transformation $\Sigma_{QQ}\rightarrow e^{2q_Q\alpha}\Sigma_{QQ}, \Sigma_{\chi\chi}\rightarrow e^{2q_\chi\alpha}\Sigma_{\chi\chi}$ is $j^\mu \propto \partial^\mu \left(f_Q q_Q \sigma_Q+f_\chi q_\chi \sigma_\chi\right)$, such that the canonically normalized pNGB corresponding to this $U(1)$ and its orthogonal combination are
\begin{align}
\sigma & = \cos \phi \sigma_Q + \sin \phi \sigma_\chi \\
\sigma' & = - \sin \phi \sigma_Q + \cos \phi \sigma_\chi 
\end{align}
with $\tan\phi = f_\chi q_\chi / f_Q q_Q$.\\
$\sigma$ is $Sp(2N_c)$ anomaly-free when $q_Q=- 3(N_c-1)q_\chi$ and thus remains a pNGB. $\sigma'$ obtains a mass from the anomaly and from $Sp(2N_c)$ instanton effects. Table \ref{tab:3} shows the anomalous couplings of $\sigma$ and $\sigma'$.

\begin{table}[tb]
\begin{center}
\begin{tabular}{c|c|c}
& $\sigma$ & $\sigma'$ \\\hline
$\kappa_g/f_\sigma$  &  $(2N_c+1)(N_c-1)/f_\sigma $                    & $(- 3(N_c-1)^2(2N_c+1)\frac{f_Q}{f_\chi})/f_\sigma $ \\
$\kappa_W/f_\sigma$ & $-6 N_c(N_c-1)/f_\sigma $                         & $-( 2 N_c \frac{f_\chi}{f_Q})/f_\sigma$ \\
$\kappa_B/f_\sigma$  & $\left[\frac{8}{3} (2 N_c+1)(N_c-1)\right. $                         & $\left[ - 8(N_c-1)^2(2N_c+1) \frac{f_Q}{f_\chi}\right. $  \\
                                    & $\left. -6 N_c(N_c-1) \right]/f_\sigma$ & $\left.  -2N_c \frac{f_\chi}{f_Q}\right]/f_\sigma$
\end{tabular}
\end{center}
\caption{Couplings of the $Sp(2N_c)$ anomaly free scalar $\sigma$ and the orthogonal $\sigma'$, where $f_\sigma\equiv \sqrt{9(N_c-1)^2 f_Q^2+f_\chi^2}$.}
\label{tab:3}
\end{table}

The neutral scalar sector of this model thus contains two pNGBs, $\eta$ and $\sigma$, and a heavy resonance $\sigma'$. A comprehensive study of the scalar sector and the general bounds on $M_\eta,M_\sigma,M_{\sigma'}, f_\chi, f_Q$ and $N_c$ from di-boson searches is under way, but beyond the scope of this letter. Here, we only investigate one special case: Can one of these states be a viable candidate for the di-boson excess reported by ATLAS \cite{Aad:2015owa} and CMS \cite{Khachatryan:2014hpa}?  $\eta$ does not couple to gluons and therefore has a too small production cross section. $\sigma$ can be made massive by explicit breaking of $U(1)_\sigma$, \ie \, external to the $SU(6)\times SU(4)\rightarrow SO(6)\times Sp(4)$ breaking. The $\sigma$ particle can be produced from gluon fusion and has decay channels into $WW$ and $ZZ$. $\sigma'$ is massive even without an explicit breaking term, and it has the required types of couplings. For $\sigma$ and $\sigma'$ we therefore inspect production and branching ratios in more detail.
\bigskip

\section{Branching ratios and production}

Ref.~\cite{Cacciapaglia:2015nga} discussed pseudo-scalars with anomalous couplings to SM gauge bosons as candidates for a di-boson excess. The state $\sigma$ falls precisely into this class, such that we can use the effective field theory analysis presented, there.

The partial widths of $\sigma$ decaying into $gg$, $WW$, $ZZ$, $Z\gamma$, and $\gamma\gamma$ are \cite{Cacciapaglia:2015nga}
\begin{align}
&\Gamma(\sigma \to gg) = \frac{g_3^4 (\kappa^\sigma_g)^2 M_\sigma^3}{128 f_{\sigma}^2 \pi^5 } 
\label{eqg1} \\
&\Gamma(\sigma \to WW) =
\frac{g_2^4 (\kappa^\sigma_W)^2 (M_\sigma^2 -4 M_W^2)^{\frac{3}{2}}}{512 f_{\sigma}^2 \pi^5 } 
\label{eqg2}\\
&\Gamma(\sigma \to ZZ) =
\frac{g_2^4 c_W^4 (\kappa^\sigma_W + \kappa^\sigma_B t_W^2 )^2 (M_\sigma^2 -4 M_Z^2)^{\frac{3}{2}}}{1024 f_{\sigma}^2 \pi^5 } 
\label{eqg3}\\
&\Gamma(\sigma \to Z\gamma) =
\frac{e^2 g_2^2 c_W^2 (\kappa^\sigma_W - \kappa_B^\sigma t_W^2)^2 (M_\sigma^2 - M_Z^2)^{3}}{512 f_{\sigma}^2 \pi^5 M_\sigma^3} 
\label{eqg4}\\
&\Gamma(\sigma \to \gamma \gamma) =
\frac{e^4  (\kappa^\sigma_W+ \kappa_B^\sigma)^2 M_\sigma^3}{1024 f_{\sigma}^2 \pi^5} 
\label{eqg5}
\end{align}
where $c_W \equiv \cos \theta_W$, $t_W \equiv \tan\theta_W$, 
$e=g_2 \sin \theta_W$ with $\theta_W$ being the weak mixing angle. With the coefficients $\kappa^\sigma/f_\sigma$ from Table~\ref{tab:3}, the branching ratios of $\sigma$ are fully determined in terms of group theoretic factors. Their range lies between
\bea
\frac{\Gamma_{\sigma\to gg}}{\Gamma_{\sigma \to WW}}&=& 5.1 \,\,\, ...\,\,\, 3.3,\\
\frac{\Gamma_{\sigma\to ZZ}}{\Gamma_{\sigma \to WW}}&=& 0.29 \,\,\, ...\,\,\, 0.31,\\
\frac{\Gamma_{\sigma\to Z\gamma}}{\Gamma_{\sigma \to WW}}&=& 0.19  \,\,\, ...\,\,\, 0.17,\\
\frac{\Gamma_{\sigma\to \gamma\gamma}}{\Gamma_{\sigma \to WW}}&=& 0.021 \,\,\, ...\,\,\, 0.033,
\eea
 where the first value is for $N_c=2$ while the last value gives the large $N_c$ limit, and we used $g_3 = 1.033$, $g_2 = 0.628$, and $\sin^2_W= 0.2319$ at
an energy of 2 TeV.

Furthermore, Ref.~\cite{Cacciapaglia:2015nga} obtained for the production cross section of a scalar resonance at LHC  run I
\beq
\sigma_I (gg\rightarrow \sigma)=\left(\frac{\kappa^{\sigma}_g}{2}\right)^2\frac{(1 \mbox{TeV})^2}{f^2_{\sigma} }0.615 \mbox{fb}.
\label{eq:prod}
\eeq
Using a Monte Carlo simulation, we determine the production cross section at LHC run II (13 TeV):
\beq
\sigma_{II} (gg\rightarrow \sigma)=\left(\frac{\kappa^{\sigma}_g}{2}\right)^2\frac{(1 \mbox{TeV})^2}{f^2_{\sigma} } 8.11 \mbox{fb}.
\label{eq:prod}
\eeq
\bigskip

\section{Comparison to experimental constraints}

The branching ratios can be compared to the experimental bounds on di-boson resonances at 2 TeV with various decay channels of the electroweak gauge bosons. The di-boson excess Ref.~\cite{Aad:2015owa} is consistent with $\sigma(gg\to \sigma)\times BR(\sigma\to WW+ZZ)\sim 10$~fb, while the resonant di-photon search \cite{Aad:2015mna} for Kaluza-Klein gravitons constrains $\sigma(gg\to\sigma)\times BR(\sigma \to \gamma\gamma)< 0.5$~fb. The search for a $Z\gamma$ resonance with $Z\rightarrow ll$ \cite{Aad:2014fha} established a bound of  $\sigma(gg\to\sigma)\times BR(\sigma \to Z\gamma)< 3$~fb for a resonance mass at 1.6~TeV while higher masses have not been considered due to low statistics. We therefore use the bound for 1.6~TeV, here. The mono-photon search \cite{Khachatryan:2014rwa} established a bound which can be re-interpreted for the $Z\gamma$ decay channel with $Z\rightarrow \nu\nu$, leading to $\sigma(gg\to\sigma)\times BR(\sigma \to Z\gamma)\cdot A\epsilon < 1$~fb, where $A\epsilon$ is the acceptance times efficiency of the model's signal.  With $A\epsilon\sim 35\%$ (corresponding to the ADD model  value in Ref.~\cite{Khachatryan:2014rwa}), this channel yields a bound comparable to the $Z_{ll}\gamma$ search of Ref.~\cite{Aad:2014fha}. The di-jet resonance search \cite{Khachatryan:2015sja} results in a bound $\sigma(gg\to\sigma)\times BR(\sigma \to gg) < 200$~fb. A stronger indirect bound can be obtained from the fact that the di-boson resonance is narrow ($\Gamma_\sigma / M_\sigma < 0.1$ \cite{Aad:2015mna}). Eqs.(\ref{eqg1}) and (\ref{eq:prod}) imply  $\sigma(gg\to\sigma)\times BR(\sigma \to gg) \lesssim 135$~fb.

Altogether, he above bounds imply 
\bea
\frac{\Gamma_{\sigma\to gg}}{\Gamma_{\sigma \to WW+ZZ}}& \lesssim & 13.5\, ,\\
\frac{\Gamma_{\sigma\to \gamma\gamma}}{\Gamma_{\sigma \to WW+ZZ}}& < & 0.05\, ,\\
\frac{\Gamma_{\sigma\to Z\gamma}}{\Gamma_{\sigma \to WW+ZZ}}& < & 0.3\, ,
\eea
and are satisfied in the model under consideration. Note that the exclusion bounds on the $Z\gamma$ and the $\gamma\gamma$ channel are not far off the predicted values, showing that di-photon, mono-photon, and $Z_{ll}\gamma$ resonance searches are very promising to test this model.

The analogous analysis for $\sigma'$ yields $\Gamma_{\sigma'\to \gamma\gamma}/\Gamma_{\sigma' \to WW+ZZ }>  0.1$ for all values of $N_c, f_Q, f_\chi$, which is in contradiction with the di-photon resonance search and thus excludes $\sigma'$ as a candidate for the 2 TeV di-boson excess. 
 
\bigskip

With the bounds on branching fractions satisfied for $\sigma$, we turn to the question whether this model can provide a sufficient production cross section for the di-boson resonance. From \refeq{eq:prod} it is clear that the production cross section can be raised by either increasing $\kappa^\sigma_g$  (which with the expression given in Table \ref{tab:3} means increasing $N_c$) or by decreasing $f_\sigma\equiv \sqrt{9(N_c-1)^2 f_Q^2+f_\chi^2}$. The branching ratios following from Eqs.(\ref{eqg1}-\ref{eqg5}) are independent of $f_\sigma$ and only mildly depend on $N_c$. 
\begin{figure}[tb]
\includegraphics[width=0.85\linewidth]{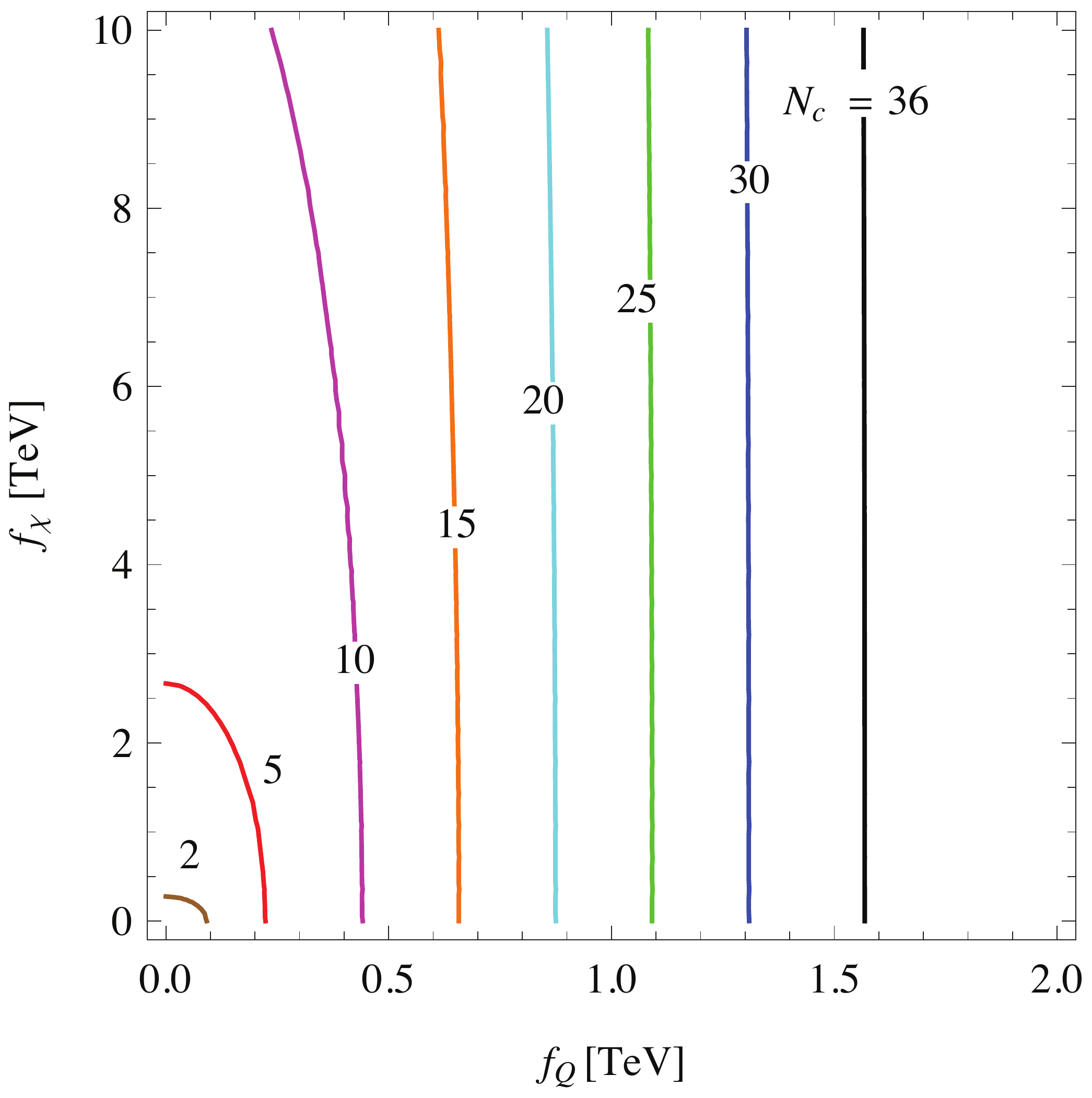}
\caption{\label{fig:fQfchi} Contours of $f_Q$ and $f_\chi$ for different $N_c$ which provide a cross section of 10 fb for $pp\rightarrow \sigma \rightarrow WW$ at LHC run I. }
\label{fig:ff}
\end{figure}

Fig.~\ref{fig:ff} shows contours for various $N_c$ in the $f_\chi$ vs. $f_Q$ parameter space for which $\sigma(gg\to\sigma)\times BR(\sigma \to WW+ ZZ) = 10$~fb. As can be seen, the cross section can be raised to 10 fb at the expense of a low $f_Q$ and/or high $N_c$. $N_c\leq36$ is required in order to maintain asymptotic freedom of the theory \cite{Ferretti:2013kya}, which from Fig.~\ref{fig:ff} implies an upper bound on $f_Q \lesssim 1.6$ TeV if $\sigma$ is supposed to be the source of the 2 TeV di-boson anomaly.

\bigskip
\section{Outlook on a 2 TeV $\sigma$ at LHC run II}

\begin{figure}[tb]
\begin{center}
\includegraphics[width=0.92\linewidth]{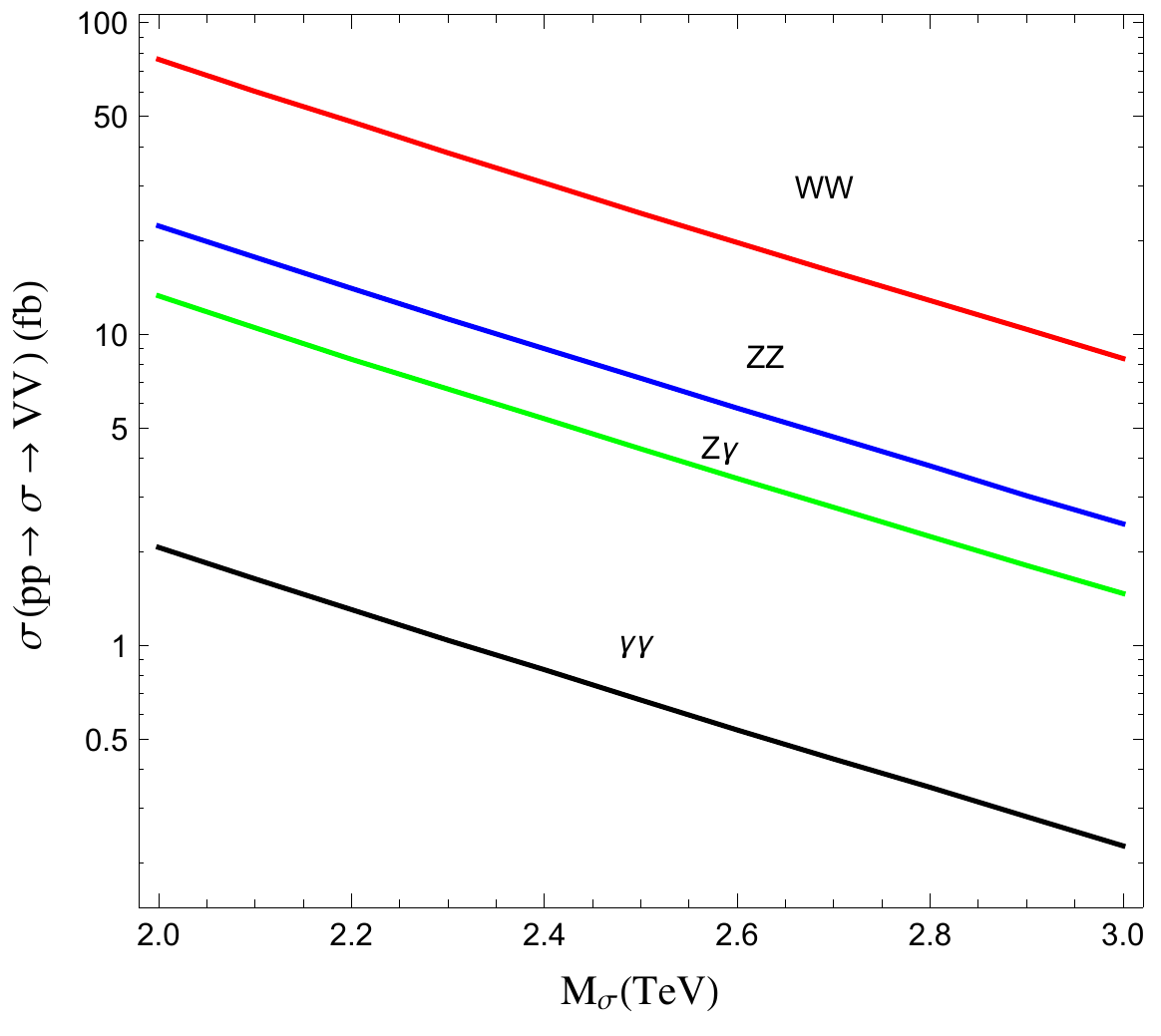}
\end{center}
\caption{Final state cross sections for the decay channels of a 2 TeV $\sigma$ resonance into different pairs of electroweak gauge bosons at LHC Run II, with $N_c =10$, $f_Q$=$f_\chi$ = 500 GeV. The result for other values can be rescaled due to the narrow width approximation.}
\label{fig:cs13TeV}
\end{figure}

\begin{figure}[tb]
\begin{center}
\includegraphics[width=0.99\linewidth]{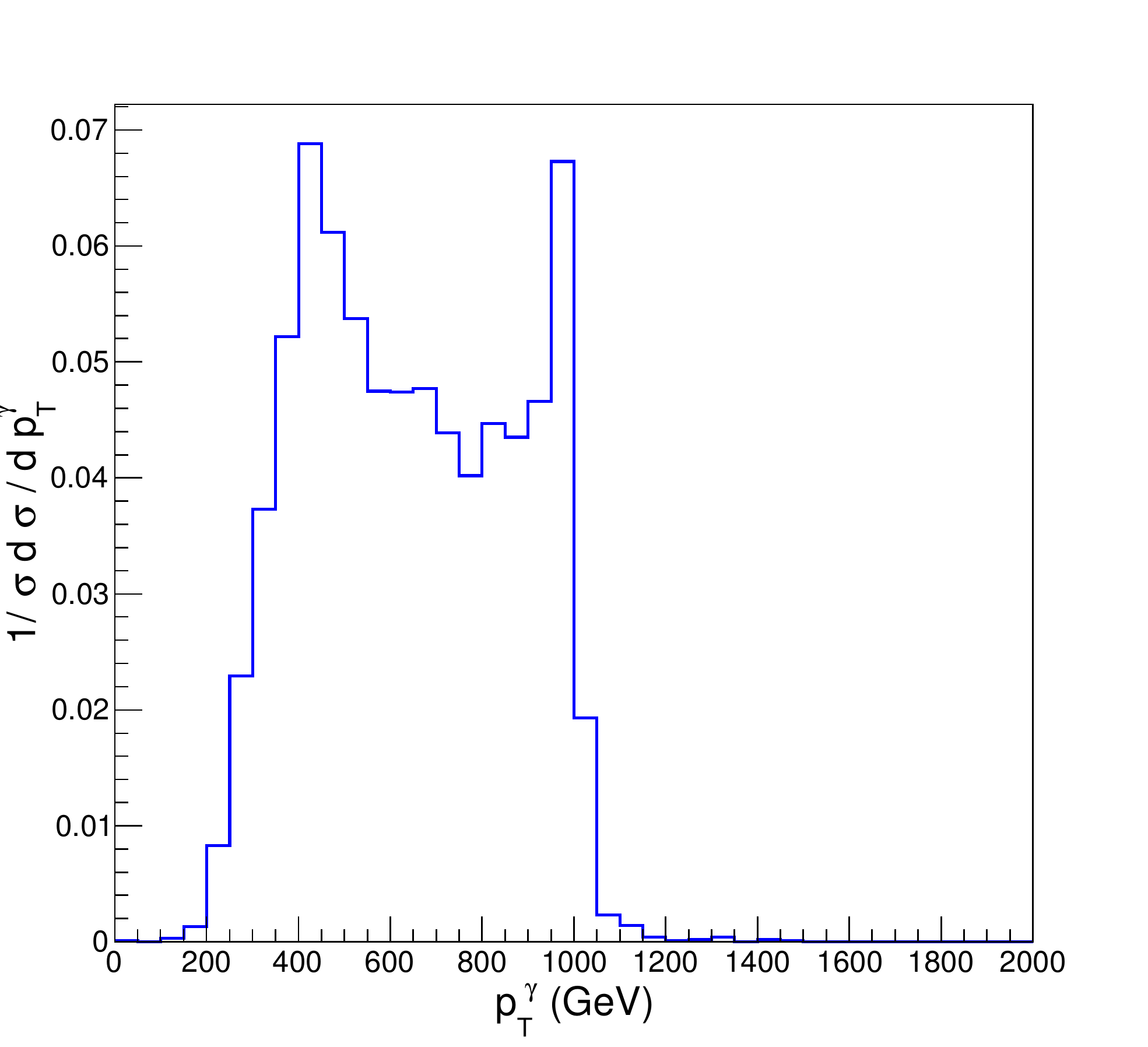}
\end{center}
\caption{$p_T$ distribution of a mono-photon occuring in the search for a 2 TeV $\sigma$ resonance in the $Z_{ll} \gamma$ channel, with $N_c =10$, $f_Q$=$f_\chi$ = 500 GeV.}
\label{fig:pta}
\end{figure}

As an outlook,  Fig.~\ref{fig:cs13TeV} shows the cross section for each channel at  $\sqrt s =$13 TeV, using parameters which generate a 2 TeV $WW + ZZ $ excess at a 8 TeV LHC.  In particular the  large signal $Z_{ll} \gamma$ signal in the high mass region,  with fully reconstructable final states, provides a golden channel to search the $\sigma$ resonance predicted in this model.  As an illustration we use Madgraph to generate  events  and perform a simple analysis for a 13 TeV  LHC, using  the same basic cut as in Ref.~\cite{Aad:2014fha}:
\bea
&&  p_T^l > 25 ~\mbox{GeV}, \quad   |\eta_l| < 2.47, \quad    65< m_{l^+ l^-} < 115~\mbox{GeV}  \nonumber  \\
&& E_T^\gamma > 40 \mbox{GeV} , \quad   |\eta_\gamma| < 2.37, \quad   \Delta R(l, \gamma) > 0.7
\eea
the $Z_{ll} \gamma$ candidate is selected by requiring two oppositely charged leptons, one isolated photon,  with their rapidity in  the detectable region of calorimeters. Specifically the requirement of  $ \Delta R(l, \gamma) > 0.7$ is used to suppress the radiation background while the mass window cut for  the $l^+l^-$ system is to ensure  the oppositely charged di-lepton originates from a decaying $Z$ boson. Since the SM background only has a  $m_{l^+l^- \gamma} $ distribution  in  the low mass region,   with adequate events we are able to  extract the mass value for a heavy  $\sigma$.  We will get $\sim 50$ events after demanding the basic cuts at a LHC Run II with $300~ \mbox{fb}^{-1}$ luminosity, for  $N_c =10$, $f_Q=f_\chi = $ 500 GeV. Thus, the $Z_{ll} \gamma$ channel is possible to test the characteristics of this model. The $p_T$ distribution is shown in Fig.\ref{fig:pta}, where the pronounced end point at $1.0$ TeV indicates a resonance with mass of  $\sim 2.0$ TeV. 

\bigskip
\section{Conclusions}
In this letter we showed that a the UV embedding of a composite Higgs model which is based on a $Sp(2N_c)$ confining gauge groups with global $SU(6)\times SU(4)\times U(1)\to SO(6)\times Sp(4)$ symmetry breaking \cite{Cacciapaglia:2015eqa, Barnard:2013zea} contains a pseudo Nambu-Goldstone boson $\sigma$ with anomalous couplings to gluons and electroweak gauge bosons. The branching ratios of $\sigma$ are fully determined by the underlying structure of the model, with the rank of the confining group being the only free parameter  which can mildly affect the branching ratios. We showed that the branching ratios allow to explain a di-boson excess in the hadronic $WW$ and $ZZ$ channel whilst being in accord with the bounds from all other LHC run I di-boson searches. The bounds of the $Z\gamma$ and the $\gamma\gamma$ searches at run I are however close to the rates predicted by the branching ratios of this model. For run II, they will form superb search channels in order to test whether a 2 TeV $\sigma$ resonance is the origin of the 2 TeV  di-boson excess of Refs.\cite{Aad:2015owa,Khachatryan:2014hpa}. Note that the $Z\gamma$ and $\gamma\gamma$ channels are generically very well suited in order to distinguish different candidate models for a di-boson resonance. For vector resonances, a decay into $Z\gamma$ or $\gamma\gamma$ is absent. For a pseudo scalar as considered here, the rates are related by group theoretic factors. Signals in these channels (and their rates) would provide a strong hint for a new scalar and its origin.
 
While the branching ratios in this model are fully fixed (apart from a mild $N_c$ dependence), the production cross section varies with the model parameters, and is thus not a prediction of the model. In our case, the model parameters are $N_c$ ,$ f_Q$ and $f_\chi$. In this letter, we show the parameter set required to reproduce a 2 TeV resonance with 10 fb production cross section. Reaching this cross section requires a high rank of the confining $Sp(2N_c)$ gauge group and a low decay constant $f_Q$ as is shown in Fig.~\ref{fig:ff}.  Note that $f_Q$ (related to the U(1) breaking) can a priori be lower than the scale $f$ (related to $SU(4)\to Sp(4)$ breaking) which governs the composite Higgs. A lower rank of the gauge group or a larger $f_Q$ would lead to a lower production cross section. Furthermore, the mass of the resonance is not fixed within the composite model, but arises from explicit $U(1)$ breaking. Thus, the resonance cross section and resonance mass are input parameters of the model (which here we chose to match the diboson resonance), while the branching ratios are predictions which can be verified in the $WW$ and $ZZ$ channel, in particular in $Z\gamma$ and $\gamma\gamma$ searches.  

More commonly considered values for the rank of the gauge group, the decay constant, and the mass of a pseudo-Goldstone boson lead to a lighter and more weakly coupled resonance, which is still interesting as it will be tested in the hadronic and semi-leptonic $WW$ and $ZZ$ channel, but in particular also in di-photon searches, $Z\gamma$ searches, and mono-photon searches.

\bigskip
\section{Acknowledgements}

We would like to thank Seung J. Lee, Alberto Parolini and Hugo Ser\^odio for collaboration at the initial stags of this work and many useful discussions. We in particular thank Giacomo Cacciapaglia and Aldo Deandrea for their input and encouragement. 
TF thanks IPNL for their hospitality during the final stages of this work. 
We acknowledge support from the Franco-Korean Partenariat Hubert Curien (PHC) STAR 2015, project number 34299VE, and thank the France-Korea Particle Physics Lab (FKPPL) for partial support. HC acknowledges partial support from the Labex-LIO (Lyon Institute of Origins) under grant ANR-10-LABX-66 and FRAMA  (FR3127, F\'ed\'eration de Recherche ``Andr\'e Marie Amp\`ere"). 
TF was supported by the Basic Science Research Program through the National Research Foundation of Korea (NRF) funded by the ministry of Education, Science and Technology (No. 2013R1A1A1062597). 

\bibliographystyle{utphys}
\bibliography{letterbib}

\providecommand{\href}[2]{#2}\begingroup\raggedright\begin{thebibliography}{10}

\bibitem{Aad:2015owa}
{ ATLAS} Collaboration, G.~Aad {\em et~al.}, ``{Search for high-mass diboson
  resonances with boson-tagged jets in proton-proton collisions at $\sqrt{s} =
  8$ TeV with the ATLAS detector},''
\href{http://arxiv.org/abs/1506.00962}{{\ttfamily arXiv:1506.00962 [hep-ex]}}.

\bibitem{Khachatryan:2014hpa}
{ CMS} Collaboration, V.~Khachatryan {\em et~al.}, ``{Search for massive
  resonances in dijet systems containing jets tagged as W or Z boson decays in
  pp collisions at $ \sqrt{s} $ = 8 TeV},''
  \href{http://dx.doi.org/10.1007/JHEP08(2014)173}{{\em JHEP} {\bfseries 08}
  (2014) 173},
\href{http://arxiv.org/abs/1405.1994}{{\ttfamily arXiv:1405.1994 [hep-ex]}}.

\bibitem{Khachatryan:2014gha}
{ CMS} Collaboration, V.~Khachatryan {\em et~al.}, ``{Search for massive
  resonances decaying into pairs of boosted bosons in semi-leptonic final
  states at $\sqrt{s} =$ 8 TeV},''
  \href{http://dx.doi.org/10.1007/JHEP08(2014)174}{{\em JHEP} {\bfseries 08}
  (2014) 174},
\href{http://arxiv.org/abs/1405.3447}{{\ttfamily arXiv:1405.3447 [hep-ex]}}.

\bibitem{Aad:2015ufa}
{ ATLAS} Collaboration, G.~Aad {\em et~al.}, ``{Search for production of
  $WW/WZ$ resonances decaying to a lepton, neutrino and jets in $pp$ collisions
  at $\sqrt{s}=8$ TeV with the ATLAS detector},''
  \href{http://dx.doi.org/10.1140/epjc/s10052-015-3593-4,
  10.1140/epjc/s10052-015-3425-6}{{\em Eur. Phys. J.} {\bfseries C75} no.~5,
  (2015) 209}, \href{http://arxiv.org/abs/1503.04677}{{\ttfamily
  arXiv:1503.04677 [hep-ex]}}.
[Erratum: Eur. Phys. J.C75,370(2015)].

\bibitem{Dias:2015mhm}
F.~Dias, S.~Gadatsch, M.~Gouzevich, C.~Leonidopoulos, S.~Novaes, A.~Oliveira,
  M.~Pierini, and T.~Tomei, ``{Combination of Run-1 Exotic Searches in Diboson
  Final States at the LHC},''
\href{http://arxiv.org/abs/1512.03371}{{\ttfamily arXiv:1512.03371 [hep-ph]}}.

\bibitem{Cacciapaglia:2015nga}
G.~Cacciapaglia, A.~Deandrea, and M.~Hashimoto, ``{Scalar Hint from the Diboson
  Excess?},'' \href{http://dx.doi.org/10.1103/PhysRevLett.115.171802}{{\em
  Phys. Rev. Lett.} {\bfseries 115} no.~17, (2015) 171802},
\href{http://arxiv.org/abs/1507.03098}{{\ttfamily arXiv:1507.03098 [hep-ph]}}.

\bibitem{Note1}
Soon after, pseudo scalar di-boson candidates were also parameterized and
  discussed in Refs.\cite {Kim:2015vba,Fichet:2015yia}.

\bibitem{Allanach:2015hba}
B.~C. Allanach, B.~Gripaios, and D.~Sutherland, ``{Anatomy of the ATLAS diboson
  anomaly},'' \href{http://dx.doi.org/10.1103/PhysRevD.92.055003}{{\em Phys.
  Rev.} {\bfseries D92} no.~5, (2015) 055003},
\href{http://arxiv.org/abs/1507.01638}{{\ttfamily arXiv:1507.01638 [hep-ph]}}.

\bibitem{Barnard:2013zea}
J.~Barnard, T.~Gherghetta, and T.~S. Ray, ``{UV descriptions of composite Higgs
  models without elementary scalars},''
  \href{http://dx.doi.org/10.1007/JHEP02(2014)002}{{\em JHEP} {\bfseries 02}
  (2014) 002},
\href{http://arxiv.org/abs/1311.6562}{{\ttfamily arXiv:1311.6562 [hep-ph]}}.

\bibitem{Cacciapaglia:2015eqa}
G.~Cacciapaglia, H.~Cai, A.~Deandrea, T.~Flacke, S.~J. Lee, and A.~Parolini,
  ``{Composite scalars at the LHC: the Higgs, the Sextet and the Octet},''
  \href{http://dx.doi.org/10.1007/JHEP11(2015)201}{{\em JHEP} {\bfseries 11}
  (2015) 201},
\href{http://arxiv.org/abs/1507.02283}{{\ttfamily arXiv:1507.02283 [hep-ph]}}.

\bibitem{Ferretti:2013kya}
G.~Ferretti and D.~Karateev, ``{Fermionic UV completions of Composite Higgs
  models},'' \href{http://dx.doi.org/10.1007/JHEP03(2014)077}{{\em JHEP}
  {\bfseries 03} (2014) 077},
\href{http://arxiv.org/abs/1312.5330}{{\ttfamily arXiv:1312.5330 [hep-ph]}}.

\bibitem{Vecchi:2015fma}
L.~Vecchi, ``{A "dangerous irrelevant" UV-completion of the composite Higgs},''
\href{http://arxiv.org/abs/1506.00623}{{\ttfamily arXiv:1506.00623 [hep-ph]}}.

\bibitem{Aad:2015mna}
{ ATLAS} Collaboration, G.~Aad {\em et~al.}, ``{Search for high-mass diphoton
  resonances in $pp$ collisions at $\sqrt{s}=8$ TeV with the ATLAS detector},''
  \href{http://dx.doi.org/10.1103/PhysRevD.92.032004}{{\em Phys. Rev.}
  {\bfseries D92} no.~3, (2015) 032004},
\href{http://arxiv.org/abs/1504.05511}{{\ttfamily arXiv:1504.05511 [hep-ex]}}.

\bibitem{Aad:2014fha}
{ ATLAS} Collaboration, G.~Aad {\em et~al.}, ``{Search for new resonances in
  $W\gamma$ and $Z\gamma$ final states in $pp$ collisions at $\sqrt s=8$ TeV
  with the ATLAS detector},''
  \href{http://dx.doi.org/10.1016/j.physletb.2014.10.002}{{\em Phys. Lett.}
  {\bfseries B738} (2014) 428--447},
\href{http://arxiv.org/abs/1407.8150}{{\ttfamily arXiv:1407.8150 [hep-ex]}}.

\bibitem{Khachatryan:2014rwa}
{ CMS} Collaboration, V.~Khachatryan {\em et~al.}, ``{Search for new phenomena
  in monophoton final states in proton-proton collisions at $\sqrt{s}$ = 8
  TeV},''
\href{http://arxiv.org/abs/1410.8812}{{\ttfamily arXiv:1410.8812 [hep-ex]}}.

\bibitem{Khachatryan:2015sja}
{ CMS} Collaboration, V.~Khachatryan {\em et~al.}, ``{Search for resonances and
  quantum black holes using dijet mass spectra in proton-proton collisions at
  $\sqrt{s} =$ 8 TeV},''
  \href{http://dx.doi.org/10.1103/PhysRevD.91.052009}{{\em Phys. Rev.}
  {\bfseries D91} no.~5, (2015) 052009},
\href{http://arxiv.org/abs/1501.04198}{{\ttfamily arXiv:1501.04198 [hep-ex]}}.

\bibitem{Kim:2015vba}
H.~M. Lee, D.~Kim, K.~Kong, and S.~C. Park, ``{Diboson Excesses Demystified in
  Effective Field Theory Approach},''
  \href{http://dx.doi.org/10.1007/JHEP11(2015)150}{{\em JHEP} {\bfseries 11}
  (2015) 150},
\href{http://arxiv.org/abs/1507.06312}{{\ttfamily arXiv:1507.06312 [hep-ph]}}.

\bibitem{Fichet:2015yia}
S.~Fichet and G.~von Gersdorff, ``{Effective theory for neutral resonances and
  a statistical dissection of the ATLAS diboson excess},''
\href{http://arxiv.org/abs/1508.04814}{{\ttfamily arXiv:1508.04814 [hep-ph]}}.

\end{thebibliography}\endgroup

\end{document}